# RENE: A Multimodal Pre-trained Framework for Pulmonary Auscultation Sounds

Pengfei Zhang[1], Zhihang Zheng[1] , Shichen Zhang[1] , Minghao Yang[1] and Shaojun Tang[1,2*]

1. Bioscience and Biomedical Engineering Thrust, Systems Hub, The Hong Kong University of Science and Technology (Guangzhou), Guangzhou, Guangdong, China.
2. Division of Emerging Interdisciplinary Areas, The Hong Kong University of Science and Technology, Clear Water Bay, Hong Kong, SAR, China.

*E-mail: Correspondence: shaojuntang@ust.hk

## ABSTRACT

Compared with invasive examinations that require tissue sampling, respiratory sound testing is a non-invasive examination method that is safer and easier for patients to accept. In this study, we introduce Rene, a pioneering large-scale model tailored for respiratory sound recognition. Rene has been rigorously fine-tuned with an extensive dataset featuring a broad array of respiratory audio samples, targeting disease detection, sound pattern classification, and event identification. Our innovative approach applies a pre-trained speech recognition model to process respiratory sounds, augmented with patient medical records. The resulting multi-modal deep-learning framework addresses interpretability and real-time diagnostic challenges that have hindered previous respiratory-focused models. Benchmark comparisons reveal that Rene significantly outperforms existing models, achieving improvements of 10.27%, 16.15%, 15.29%, and 18.90% in respiratory event detection and audio classification on the SPRSound database. Disease prediction accuracy on the ICBHI database improved by 23% over the baseline in both mean average and harmonic scores. Moreover, we have developed a real-time respiratory sound discrimination system utilizing the Rene architecture. Employing state-of-the-art Edge AI technology, this system enables rapid and accurate responses for respiratory sound auscultation(https://github.com/zpforlove/Rene).

# 1. INTRODUCTION

The global prevalence of pulmonary diseases, especially the outbreak of COVID-19, highlights the growing importance of timely detection and effective diagnosis of pneumonia. This study systematically analyzes pulmonary audio signals using audio processing techniques to achieve robust pneumonia diagnosis. Pulmonary audio signals contain valuable physiological information, and their analysis can help identify pathological manifestations of pneumonia. For example, research has shown that patients in early-stage pneumonia exhibit increased respiratory audio frequencies, concentrating energy in the high-frequency range[1]. This indicates airway narrowing, obstruction, and the presence of pulmonary inflammation, such as respiratory movement, airway constriction, and pulmonary effusion.

Analyzing these audio-derived physiological features enables a deeper understanding of pneumonia's pathological mechanisms, establishing a causal relationship between pulmonary audio signal characteristics and the disease. Furthermore, the deconvolution of audio data provides a theoretical foundation for pneumonia diagnosis and treatment. By leveraging various audio processing techniques, we can process pulmonary audio signals, extract embedded information, and develop reliable and efficient diagnostic tools and therapeutic devices for improved pneumonia diagnosis and treatment.

The fundamental principle of personalized lung sound assessment involves auscultation with a stethoscope placed on the chest. Normal lung auscultation produces two types of sounds: breath sounds and adventitious sounds[2]. Breath sounds originate from airflow during respiration, while lung resonant sounds arise from gas resonance within the alveoli and bronchi. Generally, tracheal breath sounds are created by airflow passing through the trachea, yielding louder and clearer sounds. Conversely, bronchial breath sounds occur when airflow is obstructed in the bronchi, leading to weaker and less distinct sounds[3].

Doctors can use a stethoscope's auscultation head to listen to various lung sounds and identify potential abnormalities. For example, inflammation or congestion in the lungs may produce crackles, which are typically moist sounds caused by fluid accumulation in the alveoli and bronchi, obstructing normal airflow. Conversely, the presence of a tumor or foreign object in the airways can generate loud and harsh sounds[4].

With the rapid advancement of artificial intelligence (AI) technology, there is growing interest in using deep learning for pneumonia detection and diagnosis based on audio data and medical imaging. Asif et al. developed a lightweight convolutional neural network (CNN) architecture using chest X-ray images from 2,541 confirmed COVID-19 cases and healthy controls across two public databases. Their model reliably detected COVID-19-infected patients through chest X-ray images[5]. Jain et al. compared the accuracy of Inception V3, Xception, and ResNeXt models in chest X-ray scans of COVID-19 patients and healthy individuals using the posteroanterior (PA) view, systematically evaluating their performance and accuracy[6]. Meanwhile, Mahmud proposed an effective solution for detecting COVID-19 cases even with small available X-ray sizes by efficiently utilizing trained parameters[7].

As lung sound discrimination is a commonly used non-invasive method for pulmonary examination, audio processing techniques have been widely applied in pulmonary disease analysis. For instance, Imran's research demonstrated that machine learning algorithms can analyze COVID-19 pneumonia patients' cough sounds, enabling rapid detection and diagnosis[8]. Chang et al. designed a simple homemade stethoscope to digitize medical sound data and analyzed the sounds to provide preliminary diagnoses and track health conditions[9]. In the future, audio processing techniques are expected to play an increasingly important role in pneumonia detection and diagnosis.

The motivation for designing a multi-modal system stems from the understanding that relying solely on a simple model for disease diagnosis in a clinical context is insufficient. By leveraging multiple information sources, such as lung sounds and medical record data, diagnoses become more robust, rational, and interpretable. While lung sounds offer valuable insights into respiratory system functionality and disease states, relying exclusively on them may be limited due to factors like lung anatomy and respiratory strength. Integrating lung sounds with complementary information allows for a comprehensive diagnostic assessment, facilitating a holistic understanding of the patient's condition.

In this paper, we present Rene, a multi-modal lung sound analysis architecture named after Rene Laennec, the inventor of the stethoscope. We integrated two public databases across three dimensions: respiratory event detection, respiratory recording classification, and respiratory disease identification. Notably, our study represents the first instance of fine-tuning a large-scale, general-purpose speech recognition model to create a clinically viable model for respiratory sound analysis. Finally, we describe the development of a real-time sound processing system for auscultation flow audio using a dual-thread approach. This system is trained on a combined database using the Rene architecture and deployed on the auscultation device terminal, representing an edge artificial intelligence solution for monitoring respiratory system diseases.

# 2. METHODS

## 2.1 Fine-tuning dataset

In this study, we evaluated the performance of the backbone of our proposed Rene architecture for lung sound detection using the open-source respiratory sound databases: SPRSound[10]. SPRSound, developed by Shanghai Jiao Tong University, is the first open-source pediatric respiratory sound database, containing 2,683 recordings and 9,089 events from 292 participants. The creation process involved custom sound annotation software for editing and quality assessments, with 11 experienced pediatric doctors establishing the gold standard reference.

On the other hand, ICBHI[11] is a large respiratory dataset with 920 samples from 126 patients, totaling 5.5 hours of audio and featuring 8,877 annotated crackles and 1,898 annotated wheezes. Our study focuses on the predictive accuracy of classifying eight distinct diseases using this audio data. The dataset is enriched with supplementary annotation files that supply vital metadata such as subject age, gender, and Body Mass Index (BMI), along with annotations for each respiratory cycle, specifying the presence or absence of crackles and wheezes. Furthermore, each recording in the ICBHI dataset is linked to the patient's disease condition, which is organized into three primary categories: chronic diseases (e.g., Chronic Obstructive Pulmonary Disease, Bronchiectasis, and Asthma), non-chronic diseases (e.g., upper and lower respiratory tract infections, pneumonia, and bronchiolitis), and a healthy group. This comprehensive dataset enables an in-depth investigation and understanding of disease classification, paving the way for potential advances in diagnosis and treatment.

Our experiment focuses on respiratory sound from three distinct dimensions: events, recordings, and patients. Lung sounds are categorized based on various characteristics, including duration and pitch[12]. Notably, the unclassified category can be further subdivided into continuous and discontinuous events, based on their continuity/duration. We have merged the SPRSound and ICBHI databases, meticulously curating and consolidating the data according to three task levels. This effort resulted in the creation of an integrated dataset, SFTData, comprising 16,300 respiratory sound recordings for training and 7,979 for testing.

## 2.2 The pipeline of Rene architecture

Fine-tuning the Whisper[13] speech recognition model to process respiratory sounds is a meticulous task that capitalizes on the model's versatility in handling the subtleties of breath audio. Depending on whether the priority is performance or efficiency, we tailor the model using either Lora[14] or AdaLora[15] configurations, fine-tuning parameters to resonate with the distinct patterns of respiratory acoustics. We carefully manage hyperparameters like learning rate, batch size, and number of training epochs to foster a steady and impactful learning trajectory. After performance evaluation on the test dataset, we retained the fine-tuning model, which is Whisper-small-finetune, as the best performing model. Table 1 shows the changes in Word Error Rate standards between the fine-tuning model and the original model.

**Table 1. Word Error Rate (WER) on the Test Dataset**

| Whisper Model | WER (lower is better) | Finetuned WER |
|---|---|---|
| **openai/whisper-tiny** | 140.966 | 100.999 |
| **openai/whisper-base** | 148.329 | 48.876 |
| **openai/whisper-small** | 127.632 | 25.312 |
| **openai/whisper-medium** | 139.015 | 102.093 |

While fine-tuning large-scale universal models yields satisfactory outcomes, their outputs lack interpretability, complicating the clinical analysis of respiratory diseases. To mitigate this issue and reduce associated training costs, we are investigating multi-modal transfer learning as an effective strategy. Our proposed Rene architecture is centered around the idea of multi-modal fusion diagnosis, with the flow chart illustrated in Figure 1A. This architecture comprises two primary modules, each dedicated to handling specific types of patient data to enhance the diagnostic process.

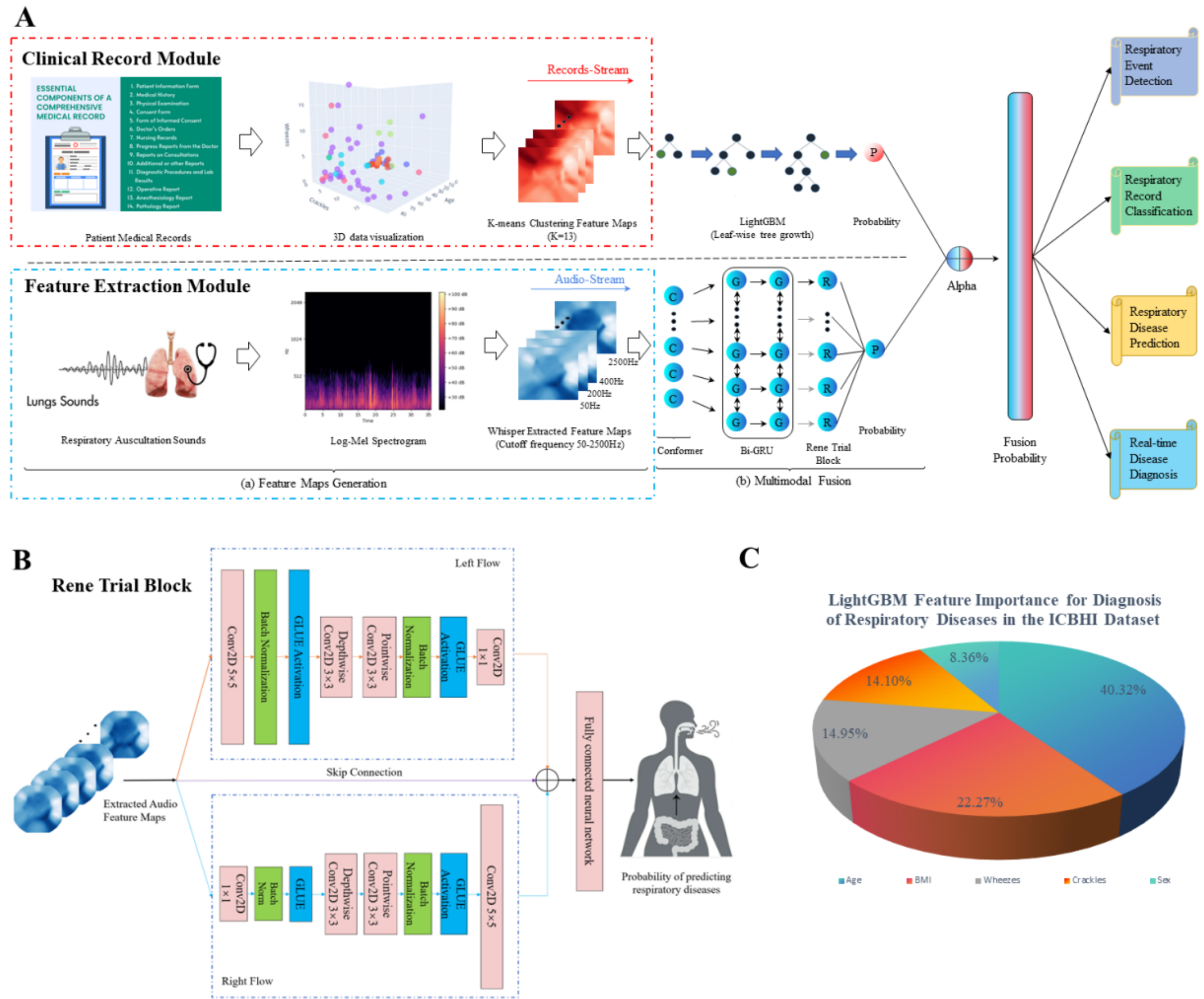

**Fig1. (A) Overview of Rene architecture.** Clinical Record Module shows the steps of Rene architecture for feature map extraction of structured information in medical records. Feature Extraction Module demonstrates the operation method of the Rene architecture for lung sound extraction feature maps. **(B) The structure of Rene Trial Block.** The triadic branch jump design progressively constricts the left bypass receptive field while broadening the right flow. Integrating residual connections within the central pathway facilitates the preservation of original features. This enhances the network's learning process, fostering stability and smoothness. **(C) Comparison of information gain for Diagnosis of Respiratory Diseases.** We employed the LightGBM algorithm to calculate the gain importance of the electronic medical records from the ICBHI Respiratory Sound Database. This quantification revealed how specific features enhance model accuracy when used for segmentation. Consequently, we could discern the relevance of key features in predicting respiratory diseases, thereby amplifying the model's interpretability and predictive precision.

Clinical Record Module is designed to extract features from patients' structured medical record data. It achieves this by employing 3D data visualization, which allow for the observation of the distribution of various feature groups within the dataset. This comprehensive visualization enables the identification of patterns and correlations that might otherwise be difficult to discern. Following this step, the K-means clustering algorithm[16] is applied to extract intra-group features, effectively segmenting the data into meaningful clusters. This procedure culminates in the creation of an electronic medical record (EMR) feature map, which serves as a valuable diagnostic tool and reference point.

We utilized the LightGBM[17] algorithm, a gradient boosting framework that employs decision tree-based learning algorithms, to gauge the gain importance of EMR from the ICBHI 2017 Challenge Respiratory Sound Database. This model allows for a structured representation of the relationships between various features and facilitates a more in-depth understanding of the complex interconnections within EMR. This data-driven approach allowed us to pinpoint specific features that substantially enhance the model's accuracy when deployed for classification tasks.

Feature Extraction Module is responsible for preprocessing of lung sound data. To accomplish this task, we utilize Whisper, a large-scale weakly supervised pre-training model. Whisper has been developed specifically to facilitate robust speech recognition and feature extraction from audio data. By applying this cutting-edge technology, we can accurately analyze patients' breath sounds and extract relevant features indicative of respiratory conditions. This analysis ultimately results in the construction of an audio stream feature map, which complements the EMR feature map generated by Clinical Record Module.

For audio stream feature information, we applied a novel network architecture, Conformer[18], which synergistically combines Convolutional Neural Networks (CNN)[19] and Transformer[20] models to extract rich and informative features from audio data. This hybrid approach capitalizes on the strengths of each model, resulting in a more robust and effective feature extraction process.

Specifically, CNN is employed for short-term temporal information extraction, enabling the network to capture local features in the audio more effectively. CNN is particularly adept at identifying and characterizing patterns within the data, making them well-suited for this task. By leveraging the power of CNN for short-term temporal analysis, we ensure that our network can accurately discern subtle variations in breath sound patterns, which are often indicative of respiratory conditions.

Meanwhile, Bidirectional Gated Recurrent Units (Bi-GRU)[21] models are utilized for long-term temporal information extraction, capturing global information within an extended range. These models are designed to handle sequential data more effectively, making them ideal for analyzing the broader patterns and trends that emerge over time in audio data. By incorporating both Transformer and Bi-GRU models into our network architecture, we enhance our ability to identify and interpret significant long-term trends in patients' breath sounds.

By integrating these complementary models, our network architecture effectively captures both local and global features within audio data, resulting in a more accurate and comprehensive representation of patients' respiratory health. This innovative approach to feature extraction from audio stream data, when combined with the insights gleaned from structured EMR, allows for a more holistic and informed diagnosis of patients' conditions, ultimately leading to improved patient care and outcomes.

Physicians often rely on the combination of multiple disease symptoms to make accurate diagnoses. Drawing inspiration from this practice, we integrate features from both audio data and EMR to create a more comprehensive diagnostic model. As illustrated in Figure 1A, patient information features are extracted from two distinct modules, with each module outputting a probability prediction value for the same task. To fuse these predictions effectively, we use the $Alpha$ confidence level, which performs a probabilistic fusion of the two modes and generates the final prediction result for our architecture. By combining the insights gained from these two modules, the Rene architecture enables a holistic approach to patient diagnosis.

In this experiment, the Rene Trial Block shown in Figure 1B employ a ternary branch jump design that gradually contracts the receptive field on the left side while expanding the right. Incorporating residual connections within the central pathway aids in preserving the original features, thereby boosting the network's learning process and promoting stability and fluency.

As shown in Figure 1C, this analysis provided valuable insights into the significance of these key features in predicting respiratory diseases. By identifying the most influential factors, the interpretability of our model was greatly amplified. This not only increased transparency but also bolstered the model's predictive precision, thereby improving the clinical applicability of our findings.

## 2.3 Insights into the architecture of Rene

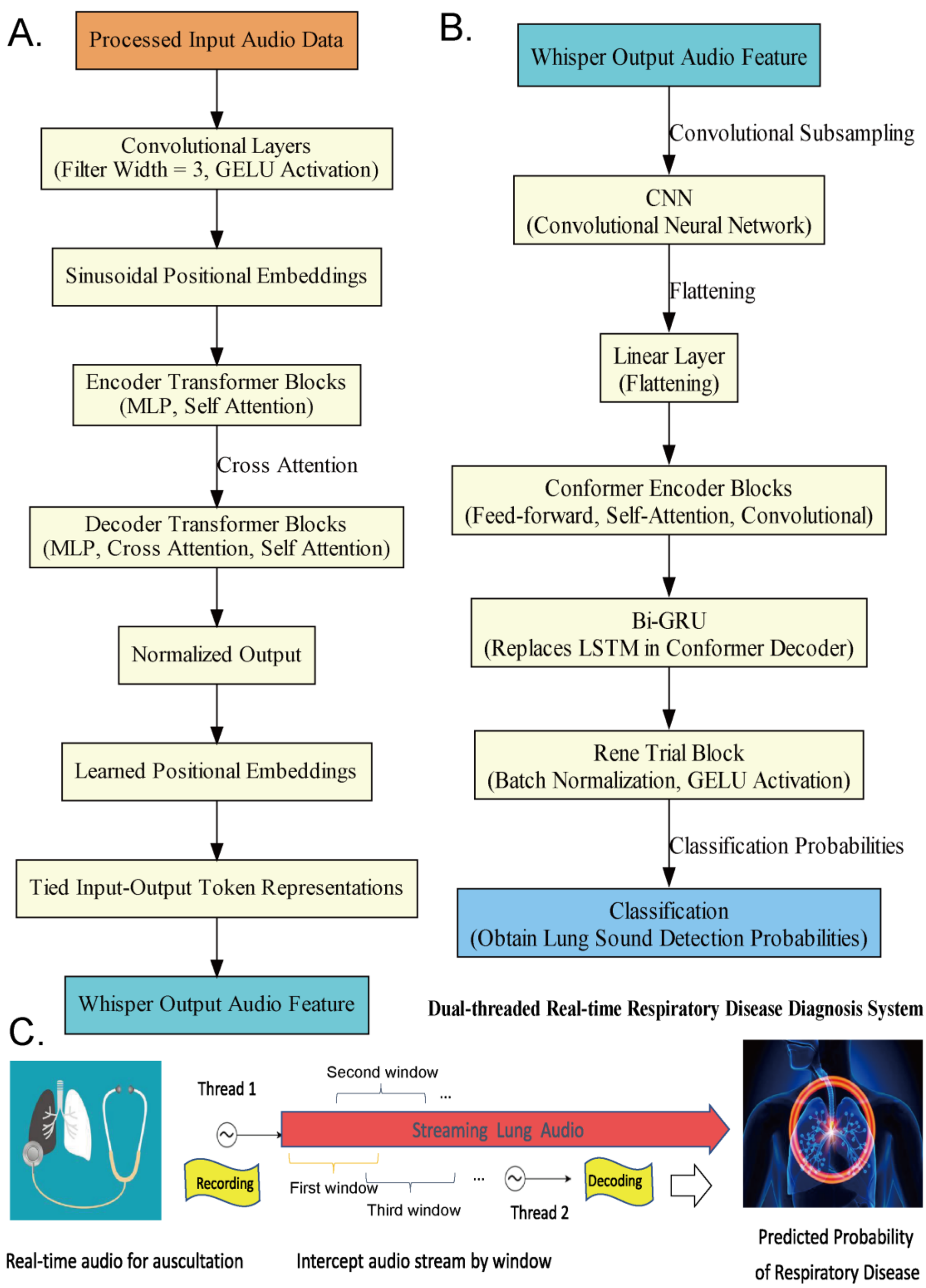

**Fig2. The key parts in Rene architecture. (A)** Whisper's Network Architecture. The Rene architecture involves the use of Whisper, a large-scale weakly supervised pre-training model specifically designed for robust speech recognition and feature extraction from audio data. Whisper's network architecture effectively processes and analyzes patients' breath sounds, identifying meaningful patterns and features that can aid in the diagnostic process. **(B)** Conformer Network Architecture. The Conformer is an advanced model that combines the strengths of Convolutional Neural Networks (CNN) and Transformer models to facilitate more effective speech recognition. By incorporating convolutional layers into the Transformer architecture, the Conformer is capable of capturing both local and global patterns in audio data, resulting in a richer representation of patients' respiratory health and ultimately improving diagnostic accuracy. **(C)** To facilitate the deployment of this model on wearable respiratory sound detection terminal devices, we have open-sourced the code and model for the dual-thread real-time respiratory disease diagnosis system at https://github.com/zpforlove/Rene. This effort aims to advance the application of edge AI in diagnosing respiratory system diseases.

To improve the accuracy and robustness of our model in analyzing pulmonary audio data, we employ several data preprocessing steps. First, the audio data is converted to numeric input and normalized and globally scaled to a range of -1 to 1, setting the pre-training dataset's mean to zero. Next, we use a 25-millisecond window with a 10-millisecond stride to compute the logarithmic mel-spectrogram representation using 80 channels.

The logarithmic mel-spectrogram offers several advantages, such as reducing audio data dimensionality, extracting discriminative and representative features, and enhancing the model's efficiency and precision. Notably, it also demonstrates favorable human perceptual attributes and resistance to interference, enabling better representation of lung sound data and potentially improved diagnostic capabilities for pulmonary diseases. The process of extracting logarithmic mel-spectrogram features from audio typically involves the following steps:

Preemphasis: Preemphasis is applied to the audio signal to boost high-frequency components. Let $x(n)$ represent the speech sample value at time n. The calculation of the pre-emphasized result $y[n]$ is as follows:

$$y[n] = x[n] - ax[n-1] \qquad (1)$$

In this case, the preemphasis coefficient $a$ is typically set to 0.97. The formula represents a first-order differencing operation, which can be interpreted as a high-pass filter.

1. Windowing: Each frame is subjected to windowing, commonly using a Hamming window. The formula for calculating the Hamming window is as follows, where $N$ denotes the order:

$$w_{ham}(n) = \alpha - \beta \cdot cos\left(\frac{2\pi n}{N-1}\right) \; \alpha = 0.53836, \beta = 0.46164 \qquad (2)$$

2. Fast Fourier Transform (FFT): After windowing each frame, a Fast Fourier Transform (FFT) is applied to obtain the frequency spectrum. This approach substantially diminishes computational complexity, rendering the FFT more efficient than direct Fourier transform calculation. The formula for the FFT calculation is as follows:

$$X(k) = \sum_{n=0}^{N-1} x[n] e^{\frac{-j2\pi kn}{N}} \quad (3)$$

In the equation, $X(k)$ represents the k-th frequency component of the frequency domain signal, $x(n)$ represents the sampled value of the time domain signal at discrete time point n, and N represents the length of the time domain signal.

3. Filtering with a set of Mel filters: After the FFT, the resulting signal is passed through a set of Mel filters to obtain the energy distribution in the Mel frequency scale. The conversion formulas between Mel scale and Hz are as follows:

$$\begin{cases} mel = 2595 log_{10}\left(1 + \frac{hz}{700}\right) \\ hz = 700\left(10^{\frac{mel}{2595}} - 1\right) \end{cases} \quad (4)$$

4. Logarithm transformation of filtered energy: After filtering the signal, we logarithmically transform the energy values to obtain Mel Frequency Cepstral Coefficients (MFCC). This step compresses the signal's dynamic range and emphasizes crucial features.

We generate the logarithmic mel-spectrogram from the audio signal and obtain the resulting MFCC from the mel-spectrogram to serve as input for the pre-trained lung sound analysis model. The logarithmic mel-spectrogram offers a compact representation of the audio signal, capturing essential frequency and temporal information for further analysis and modeling.

We employed the Whisper speech recognition model, which follows a Transformer encoder-decoder architecture. The encoder takes processed input data and applies two convolutional layers with a filter width of three and the GELU activation function[22]. It then adds sinusoidal positional embeddings to the main output, which are carried over to the encoder Transformer blocks. Pre-activation residual blocks[23] maintain the model's stability and efficiency, and the last layer of the encoder output is normalized. In the whisper decoder, the model uses learned positional embeddings and tied input-output token representations[24]. Overall, the Whisper model aims to leverage the strengths of these components for accurate and efficient speech feature processing. Figure 2A illustrates the specific structure.

The Transformer model excels at capturing global interactions, while the CNN is particularly adept at utilizing local features. In our proposed Rene architecture, we incorporate a structure known as Conformer into the backbone network, enabling it to model the semantic audio features extracted from Whisper decoding. This melding of the CNN and Transformer facilitates simultaneous consideration of both local and global dependencies within the audio features.

Initially, the audio encoder component of the Conformer applies a convolutional subsampling layer to analyze the audio features generated by the Whisper module. Following the flattening of the input data using a linear layer, multiple Conformer Blocks are employed for encoding purposes. Each of these blocks comprises two feed-forward layers featuring half-step residual connections, interspersed with multi-head self-attention and convolutional modules. Through the creation of this highly integrated architecture, we effectively harness the strengths of various models to enhance the performance of the Automatic Lung Sound Recognition task.

In the original Conformer model, a single LSTM[25] layer served as the decoder. However, a study by Fu-Shun Hsu et al. on the HF_Lung_V1 open-source lung sound database[26] revealed that Bi-GRU consistently outperformed LSTM in various detection tasks, as evidenced by significantly higher F1-scores. Consequently, this study replaces the decoder layer with Bi-GRU. RNN networks are characterized by each node receiving the output (i.e., hidden state) of the previous node as input. Accordingly, the information contained in the last node of the decoder is fed into Rene Trial Block. The Rene Trial Block processes the information to obtain classification probabilities for lung sound detection.

In medical image recognition, widely-used CNNs and other models typically treat pixel data as a single input modality, neglecting to incorporate contextual clinical information. This limitation can hinder their clinical application and translation. For instance, in a patient with fever and elevated white blood cell count, a chest X-ray examination may accurately indicate pneumonia. However, for patients lacking supporting clinical features and laboratory results, similar radiographic findings could represent alternative etiologies, such as atelectasis, pulmonary edema, or even lung cancer[27].

Similar examples can be found in the medical audio analysis domain, where clinical contextual information—typically derived from structured electronic health records data—is vital for achieving accurate and reliable results. The model architecture designed in this study employs a multi-modal approach, training separate models using features from different modalities, such as electronic medical records and audio data. The output probabilities of these multiple models are fused using confidence weighting to make the final decision. The selection of confidence weighting is often empirical and depends on the application scenario and input modalities.

## 2.4 Hyperparameter settings

In databases with complete patient information, records are typically structured data. Ravid et al. discovered that integrating tree-based model with deep learning models yielded optimal results for structured feature datasets[28]. The ultimate goal of this study is to deploy the model on edge AI devices, creating a real-time lung sound discrimination system. Consequently, there is a high demand for the model's real-time diagnostic speed. LightGBM significantly outperforms XGBoost in computational speed and memory consumption. As a result, we use LightGBM in our Rene architecture to process EMR. The prediction probability output of the LightGBM model is then fused with Rene Trial Block for the final lung sound classification task discrimination.

Most machine learning algorithms have inherent limitations in constructing individual models. The more information we gather before making a decision, the greater the likelihood of making the correct decision. Therefore, our natural inclination is to integrate predictions from multiple models to overcome the shortcomings of individual models and provide a comprehensive final decision. Given that clinical diagnosis requires the integration of patient medical records and multi-modal information such as audio auscultation, we choose to use confidence to combine the probability predictions from the output of Rene Trial Block and the classification probability values from LightGBM. The calculation method is as follows:

$$Output(p_t) = Alpha * Rene(p_t) + (1 - Alpha) * LightGBM(p_t) \quad (5)$$

Where $p_t$ represents the estimated probability of the numerical labels for each disease category. $Alpha$ is a hyperparameter that guides the prediction results to be more biased towards either audio features or medical records. To showcase the effectiveness of the multi-modal fusion Rene architecture, we conducted an ablation study to investigate the impact of the $Alpha$ value, which is chosen from the range of 0 to 1 with a step size of 0.1.

As $Alpha$ increases, the model leans more towards the output of the audio feature extraction network, leading to a progressive improvement in specificity, indicating enhanced accuracy in detecting healthy subjects. However, the model's sensitivity initially increases and then decreases, signifying a rise in the misdiagnosis rate for patients with respiratory diseases. The Rene architecture at $Alpha$=0.2 demonstrates the best performance on the ICBHI test set, with a 23% improvement in the mean of average score and harmonic score compared to the baseline, showcasing its competitive advantage.

In this study, we explored various combinations of model hyperparameters, such as network depth, model dimension, and the number of attention heads, to select the best-performing models within the constraints of model parameter size. Ultimately, we finalized two models: Rene (S), a small model with 79M parameters, and Rene (L), a large model with 1752M parameters. In our experiments, Rene (L) was utilized for model performance validation, while Rene (S) was deployed on intelligent terminal devices. Table 2 outlines the architectural hyperparameters of these models.

**Table 2. Hyperparameter of Rene (S) and Rene (L)**

| Model | Rene(S) | Rene(L) |
|---|---|---|
| **Whisper Layers** | 4 | 32 |
| **Whisper Output Dim** | 384 | 1280 |
| **Whisper Attention Heads** | 6 | 20 |
| **Conformer Encoder Layers** | 16 | 17 |
| **Conformer Encoder Dim** | 256 | 512 |
| **Conformer Attention Heads** | 4 | 8 |
| **BiGRU Decoder Dim** | 512 | 512 |

## 2.5 Real-time auscultation system

Presently, artificial intelligence (AI) computations predominantly occur in data centers or the "cloud." However, cloud-centric architectures may not be ideal for addressing information security concerns and power consumption challenges in product design. We advocate for the development of edge-based AI for diagnosing respiratory system diseases using lung sound data, wherein algorithm models are deployed on intelligent terminals. This approach offers several advantages. First, it facilitates real-time decision-making by reducing network latency and conserving bandwidth. Second, it significantly enhances data security by eliminating the need for data transmission, which is essential for safeguarding patient privacy.

To optimize the computational efficiency of edge AI devices, we have successfully compressed the parameter size of the Rene model while maintaining its prediction accuracy without a significant decrease. The smaller, distilled model allows for rapid and precise real-time identification of respiratory sound pathologies in clinical environments. We strongly recommend implementing the streamlined Rene(S) architecture to enhance diagnostic capabilities.

The complexity of data in terminal systems for diverse populations often renders a single database ineffective in capturing all common lung sound features. To overcome this, we combined the International Conference on Biomedical and Health Informatics (ICBHI) database and the King Abdulaziz University Hospital (KAUH[29]) database, resulting in 1,256 respiratory sound recordings from 238 participants for model training. The ICBHI database offers event-level annotations for respiratory sounds, comprising 3,642 Normal, 1,864 Crackles, 886 Wheezes, and 506 Crackles & Wheezes. Meanwhile, the KAUH database provides recording-level annotations, including 35 Normal, 23 Crepitus, 41 Wheeze, 8 Crackle, 1 Bronchial, 2 Wheeze & Crackle, and 2 Bronchial & Crackle. The fusion databases encompass detailed patient information, such as records of 11 common diseases (e.g., Asthma, Heart Failure, Pneumonia, Bronchitis, Bronchiectasis, Bronchiolitis, Pleural Effusion, Lung Fibrosis, Lower Respiratory Tract Infection (LRTI), Upper Respiratory Tract Infections (URTI), and Chronic Obstructive Pulmonary Disease (COPD)), along with respiratory sound data from a total of 15 diseases. By combining these databases, our data ensures adequate generalization and adaptation for real-world scenarios.

In this experiment, we implemented a real-time respiratory sound flow diagnosis system on edge AI devices using a dual-thread design. One thread handles microphone audio recording (Recording), while the other manages real-time dynamic decoding (Decoding on-the-fly). The stethoscope connected to the terminal's microphone processes data in 10ms audio data point units and feeds it into a 60-minute ring buffer. Every 10 seconds, a Mel feature calculation window extracts streaming audio data, which is then input into the model for real-time probability inference of disease classification, as depicted in Figure 2C.

In conclusion, the Dual-Thread Auscultation Stream's automatic identification solution for real-time disease probability output, based on the Rene model, holds promising prospects and practicality. It effectively improves the accuracy and efficiency of medical diagnosis. By concurrently processing audio recordings and dynamically decoding outputs, the dual-threaded real-time respiratory disease diagnosis system enhances the efficiency of the diagnostic process, enabling healthcare professionals to make more informed decisions and provide better patient care. Furthermore, the system's real-time capabilities facilitate continuous monitoring of patients' respiratory health, allowing for proactive identification and management of potential issues, ultimately contributing to improved patient outcomes.

# 3. RESULTS

## 3.1 Experimental task details

The SPRSound training set comprises 1,949 audio recordings and 6,656 respiratory sound events from 251 participants. The test set contains 734 audio recordings and 2,433 respiratory sound events, featuring both the same participants from the training set and additional individuals. Crucially, respiratory sounds are categorized into normal and adventitious sounds. Adventitious sounds can be further classified based on their duration as continuous adventitious sounds (CAS)—including Rhonchi, Wheeze, and Stridor—or discontinuous adventitious sounds (DAS), such as Coarse Crackle and Fine Crackle[30]. Given the SPRSound database, the developers proposed two levels of classification tasks to address the dataset's specific characteristics.

**Task 1 (Respiratory Sound Classification at the Event Level)**

Task 1-1: This binary classification task aims to categorize respiratory sound events as either "Normal" or "Adventitious." The goal is to distinguish normal respiratory sounds from those exhibiting adventitious characteristics.

Task 1-2: This multi-class classification task seeks to categorize respiratory sound events into seven classes, including "Normal," "Rhonchi," "Wheeze," "Stridor," "Coarse Crackle," "Fine Crackle," and "Wheeze & Crackle." The objective is to differentiate between normal respiratory sounds and various types of adventitious sounds, such as specific categories of crackles, wheezes, and other abnormal respiratory sounds.

**Task 2 (Respiratory Sound Classification at the Recording Level)**

Task 2-1: This ternary classification task aims to categorize respiratory sound recordings as "Normal," "Adventitious," or "Poor Quality." The goal is to differentiate between normal recordings, those with adventitious sounds, and poor-quality recordings that may impact the analysis's accuracy.

Task 2-2: This multi-class classification task seeks to categorize respiratory sound recordings into five classes, including "Normal," "Continuous Adventitious Sounds (CAS)," "Discontinuous Adventitious Sounds (DAS)," "CAS & DAS," or "Poor Quality." The objective is to classify recordings based on the presence of normal sounds, continuous or discontinuous adventitious sounds, and poor-quality recordings with artifacts or other issues affecting the analysis.

Sensitivity (SE) and Specificity (SP) are widely used in the medical field to evaluate the accuracy of diagnostic tests[31]. In assessing lung sound detection models, we adopt evaluation criteria from the IEEE BioCAS 2022 Respiratory Sound Classification Challenge[32]. The evaluation metrics for each task encompass Sensitivity (SE), Specificity (SP), Average Score (AS), Harmonic Score (HS), and Final Score, defined as follows:

$$SE = \frac{\sum P_i}{\sum N_i} \tag{6}$$

$$SP = \frac{P_N}{N_N} \tag{7}$$

$$AS = \frac{SE+SP}{2} \tag{8}$$

$$HS = \frac{2*SE*SP}{SE+SP} \tag{9}$$

In this paper, $P_i$ and $P_N$ represent the number of accurately classified samples for adventitious class i and Normal class, respectively. Conversely, $N_i$ and $N_N$ denote the total number of samples for adventitious class $i$ and Normal class individually. These definitions will be crucial for our analysis and discussion. The final score for each task is the mean of the Average Score (AS) and Harmonic Score (HS). This provides a concise and balanced evaluation of the model's performance.

$$Score = \frac{AS + HS}{2} \tag{10}$$

In this experiment, we employ a Mel filter bank with a cutoff frequency range of 50 Hz to 2.5 kHz for respiratory sounds, chosen to suppress background noise from equipment aging or operational errors, leaving only clean respiratory sound signals. Upon analyzing the data, we observed a significant imbalance in the SPRSound dataset, with the "Normal" class vastly outnumbering other classes at both the respiratory event and audio recording levels. Drawing inspiration from Li Jun et al.[33], we implemented two methods to address this severe class imbalance issue: WeightedRandomSampler[34] and Focal Loss[35].

Due to hardware limitations, such as the size of the CUDA memory, it may not be feasible to load a large amount of data into the network for training all at once. Therefore, it becomes necessary to read the data in batches. When encountering the problem of imbalanced sample data, WeightedRandomSampler (WRS) can be employed to balance the samples within each batch based on their respective weights, thereby improving the performance of the model.

Focal Loss is a variant of Cross Entropy Loss (CE). When an overwhelming number of easily distinguishable negative samples dominate the CE loss, focus on positive samples is diminished. Focal Loss addresses this issue by introducing a dynamic scaling factor that reduces the weight of easily distinguishable samples during training. This approach allows the model to quickly shift its focus to more challenging samples, enabling them to dominate the gradient during training. The formula for Focal Loss is as follows:

$$F_l\ (p_{\mathrm{t}}) = -(1 - p_{\mathrm{t}})^{\gamma} \log\ (p_{\mathrm{t}}) \tag{11}$$

The modulation factor $(1 - p_{\mathrm{t}})^{\gamma}$ is utilized in Focal Loss to reduce the contribution of easily classified samples. As $p_{\mathrm{t}}$ increases, indicating that a sample is easier to distinguish, the modulation factor decreases. The parameter $\gamma$ controls the rate at which the loss is reduced for easily classified samples and can take values in the range of [0, 5]. When $\gamma$ is set to 0, the Focal Loss reduces to the original Cross Entropy Loss. In the original Focal Loss paper, the authors found that $\gamma = 2$ yielded the best experimental results. With $\gamma = 2$, the loss for an example with $p_{\mathrm{t}} = 0.9$ is reduced by a factor of 100 compared to the Cross Entropy Loss, and when $p_{\mathrm{t}} \approx 0.968$, the loss is reduced by a factor of 1000. Inspired by these findings, we adopt $\gamma = 2$ as the default parameter for training the Rene architecture with Focal Loss in this study.

We trained the model using an NVIDIA Tesla A100 GPU on an Ubuntu 22.04 LTS platform. During the training process, we set the default batch size to 16, the number of epochs to 60, and the initial learning rate to 0.000001. The learning rate exponentially decayed by a factor of 0.1 every 2k steps, ensuring better stability in the later stages of training.

# 3.2 Model comparison

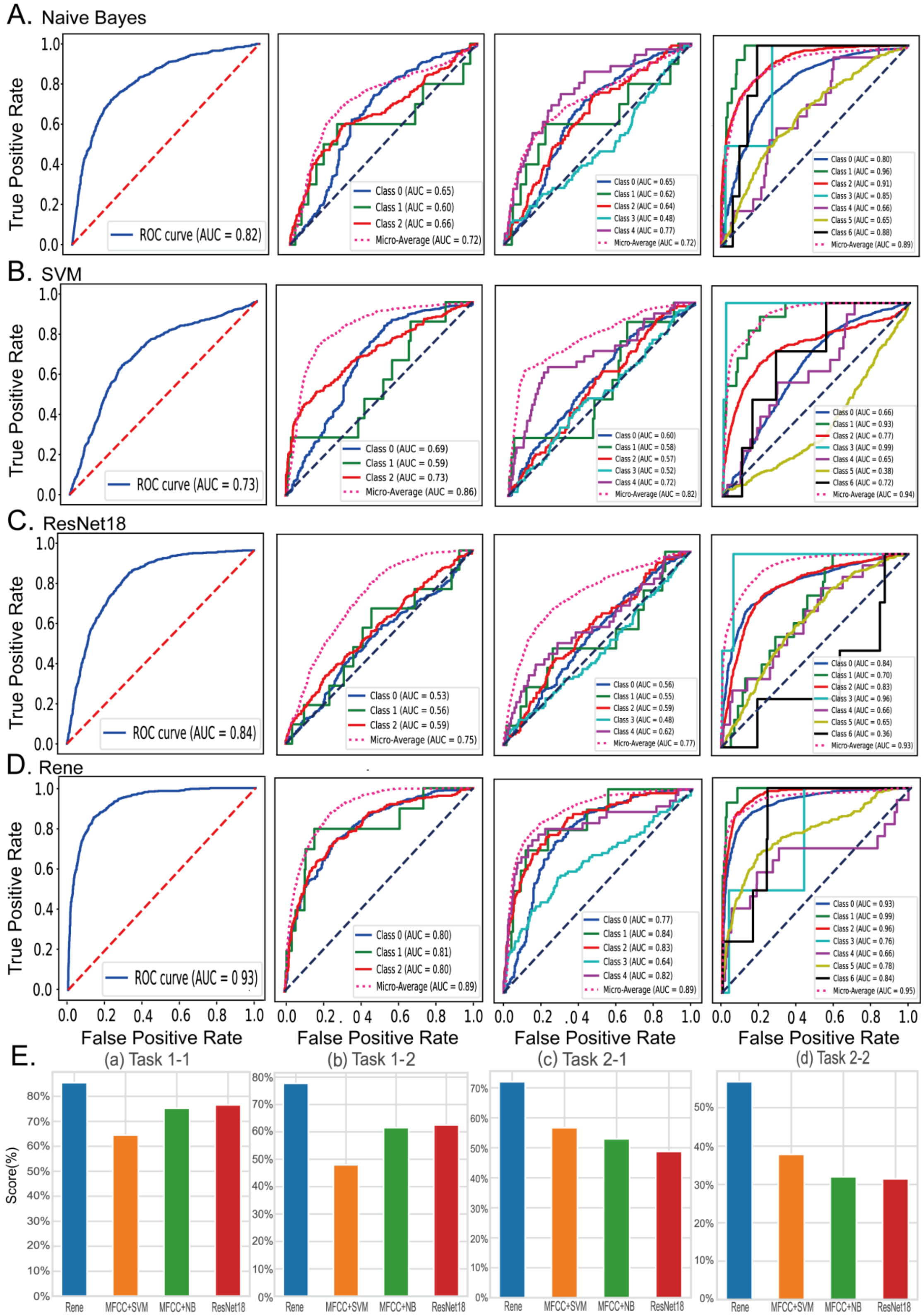

**Fig3. Performance Comparison of Various Models on the SPRSound Database.**

(A) **(B) (C) (D).** We conducted a detailed comparison of the performance of different models on the SPRSound database, encompassing Naive Bayes, SVM, ResNet18, and our proposed Rene model. The analysis comprises four distinct classification tasks, with the results displayed in panels (A), (B), (C), and (D). The panels illustrate the Area Under the Curve (AUC) and Receiver Operating Characteristic (ROC) scores for each model, providing a comprehensive assessment of their performance across the SPRSound database classification tasks. **(E).** Panel E offers a direct comparison of Rene's performance against the other models. The score enables an unbiased assessment of the relative performance of our proposed Rene architecture, highlighting its potential advantages in the realm of lung sound detection and respiratory disease diagnosis.

The mel frequency cepstral coefficient (MFCC)[36] method effectively simulates the response of the human auditory system and provides features with excellent discrimination and low correlation in various patient combinations, inter-patient comparisons, and overall classification analyses. The MFCC method, when combined with the Naive Bayes (NB)[37] classifier, demonstrated superior performance in tasks 1-1 and 1-2 compared to other machine learning algorithms. This underscores the effectiveness of low correlation MFCC features in achieving robust classification results using the NB classifier. Furthermore, the MFCC method paired with the support vector machine (SVM)[38] classifier also exhibited high classification performance in tasks 2-1 and 2-2.

We utilized the best-performing machine learning methods for each task, as provided by the SPRSound database developers, to serve as baseline models. These baselines achieved scores of 75.22%, 61.57%, 56.71%, and 37.84% in the four classification challenges at the event and recording levels, respectively[43]. In comparison, our Rene architecture demonstrated a marked improvement, attaining the highest scores of 85.49%, 77.72%, 72.00%, and 56.74% in the four tasks, respectively. When compared to the baseline models, our proposed Rene architecture improved performance by 10.27%, 16.15%, 15.29%, and 18.90% in the corresponding tasks. These results not only highlight the significant advantages of our proposed architecture for lung sound detection tasks but also underscore the potential for enhancing diagnostic accuracy in respiratory system disease diagnosis.

To further compare the performance of our Rene model with state-of-the-art lung sound detection models, we reproduced the top-performing model (with ResNet18 as the backbone) from the IEEE BioCAS 2022 Respiratory Sound Classification Challenge. We then compared its scores on the SPRSound database tasks with those of our Rene model to gain a deeper understanding of the relative performance of our proposed architecture. The results of this comparison are depicted in a bar graph in Figure 3E, which clearly demonstrates the competitive performance of our Rene architecture in lung sound detection tasks.

The increased performance of the Rene architecture can be attributed to its multi-modal fusion approach, which integrates information from both audio data and electronic medical records. This combination of data sources enables a more comprehensive understanding of patients' conditions, which in turn leads to better diagnostic performance. Furthermore, the innovative network architecture and ternary branch jump design, which effectively facilitate multi-scale feature learning, resulting in a more accurate representation of patients' respiratory health.

In conclusion, the improved performance of the Rene architecture highlights the value of employing advanced model architecture design and multi-modal fusion approaches in respiratory system disease diagnosis. By integrating information from different data sources and leveraging state-of-the-art AI models, the Rene architecture has the potential to significantly enhance the accuracy of diagnoses and ultimately improve patient care and outcomes.

# 4. CONCLUSION

In this study, we introduce a novel multi-modal architecture called Rene, which is named after Rene Laennec, the inventor of the stethoscope. The architecture employs the Whisper pre-trained model to extract decoded audio features, which are then re-encoded and re-decoded through the Conformer and BiGRU networks. We fuse the predicted disease probabilities from the Rene Trial Block with the electronic medical record feature classification probabilities computed by LightGBM, using the confidence level $Alpha$.

Compared to the baseline model of SPRSound database, the proposed Rene architecture achieves performance improvements of 10.27%, 16.15%, 15.29%, and 18.90% in four tasks, respectively. In patient disease prediction tests conducted on the ICBHI database, the mean of average score and harmonic score of the proposed model improved by 23%, compared to the baseline model. The algorithm effectively classifies lung diseases and healthy populations with high accuracy while minimizing the risk of cross-infection.

We also propose a real-time disease diagnosis system based on Rene architecture for analyzing respiratory audio data. This system features a dual-thread design that enables simultaneous microphone recording and real-time dynamic decoding through compressed model parameters. This edge-AI art facilitates rapid response disease diagnosis and can be implemented on wearable clinical detection devices, further advancing the development of edge AI in the realm of respiratory disease detection.

# 5. DISCUSSION

The proposed Rene architecture demonstrates promising performance in tasks such as respiratory sound discrimination and real-time diagnosis. We suggest three main directions for future work:

1. Integrating CT imaging data with the multi-modal Rene model for lung disease diagnosis: Current research primarily focuses on methods for lung sound data processing. Incorporating CT imaging data can offer more comprehensive lung information, thereby improving diagnosis accuracy and reliability. Future work should explore effective multi-modal fusion methods and confidence-weighted integration of CT imaging data within the Rene model to enhance its utilization and generate reliable diagnostic results.

2. Simulating lung sound data using GAN[39] models: Employing generative adversarial network (GAN) models can simulate and generate high-fidelity lung sound data, increasing the training dataset's size and diversity. This will improve the model's robustness and generalization ability, enabling it to adapt to a broader range of respiratory sound data. Combining GAN techniques with lung disease diagnosis models is crucial for enhancing model performance and reliability.

3. Synergistic adaptation with large language models (LLMs) such as GPT-4[40] in downstream tasks: LLMs like GPT-4 represent breakthroughs in artificial intelligence, transitioning from weak to general AI. Extending Rene with LLMs in downstream tasks can further expand lung disease diagnosis application domains. As real-time respiratory sound data rapidly accumulates from precision medicine, training larger, deeper models becomes possible. Edge AI models can then acquire stronger semantic understanding and reasoning capabilities, enhancing their performance in more complex tasks, such as real-time clinical decision support. This synergistic evolution will foster greater breakthroughs and advancements in lung disease diagnosis.

In conclusion, integrating CT imaging data with the multi-modal Rene architecture, simulating lung sound generation using GAN models, and synergistically evolving with large models like GPT-4 are potential future extensions of our work. Investigating these directions will contribute to improving lung disease diagnosis models' performance, reliability, and application scope, driving progress in this field.

## AUTHOR CONTRIBUTIONS

Conceptualization: Pengfei Zhang, Zhihang Zheng, Shaojun Tang. Algorithm development and implementation: Pengfei Zhang. Investigation: Pengfei Zhang, Minghao Yang. Visualization: Pengfei Zhang, Shichen Zhang. Supervision: Shaojun Tang, Pengfei Zhang. Writing—original draft: Pengfei Zhang, Shaojun Tang. Writing—review & editing: Pengfei Zhang, Shaojun Tang, Minghao Yang. All authors approved the final manuscript.

## FUNDING

We express our profound gratitude to the funding organizations that generously supported our research endeavors. Our heartfelt thanks go to the Guangzhou municipal government for bestowing the Basic Science Pilot Award 2023 upon us, significantly facilitating our initial research activities. Additionally, we appreciate the Guangdong Provincial Department of Education for honoring our work with the Education Innovation Project Award 2023. This prestigious accolade not only offered financial support, but also inspired our team to pursue innovative solutions in our field relentlessly. Lastly, we extend our sincere appreciation to the Hong Kong University of Science and Technology (HKUST) Center for Aging Science for their invaluable 2022 Seed Funding (Project number: Z1056). These supports have empowered us to explore novel avenues and broaden our research scope, striving for groundbreaking discoveries in respiratory diseases.

## CODE AND DATA AVAILABILITY

The respiratory sound data utilized in this study were sourced from publicly available and well-validated databases, ensuring repeatability and generalization. These databases can be found at the following links:

1. ICBHI Open-Source Respiratory Sound Database: This comprehensive database can be accessed at https://bhichallenge.med.auth.gr/ICBHI_2017_Challenge.

2. Shanghai Jiao Tong University Pediatric Respiratory Sound Database (SPRSound): This open-source pediatric-specific database is available at https://github.com/SJTU-YONGFU-RESEARCH-GRP/SPRSound.

3. King Abdullah University Hospital (KAUH) Database: This repository can be found at https://data.mendeley.com/datasets/jwyy9np4gv/3.

The codes, pretrained models, and relevant resources for Rene are publicly available with a detailed guide on the GitHub repository: https://github.com/zpforlove/Rene. All data needed to evaluate the conclusions in the paper are present in the paper and/or the Supplementary Materials.

## DECLARATION OF COMPETING INTREST

On behalf of all authors, the corresponding author states that there is no conflict of interest.